\documentclass[conference]{IEEEtran}
\usepackage{graphicx}  % Written by David Carlisle and Sebastian Rahtz

\usepackage{amsmath}   % From the American Mathematical Society
\usepackage{dcolumn}
\begin{document}

% paper title
\title{Stability limits of an optical frequency standard based on free Ca atoms}

% author names and affiliations
% use a multiple column layout for up to three different
% affiliations
\author{\authorblockN{J.\ A.\ Sherman and C.\ W.\ Oates \\ }
\authorblockA{NIST\\
325 Broadway\\
Boulder, CO  80027\\
Email: jeff.sherman@nist.gov}}

% avoiding spaces at the end of the author lines is not a problem with
% conference papers because we don't use \thanks or \IEEEmembership

% for over three affiliations, or if they all won't fit within the width
% of the page, use this alternative format:
% 
%\author{\authorblockN{Michael Shell\authorrefmark{1},
%Homer Simpson\authorrefmark{2},
%James Kirk\authorrefmark{3}, 
%Montgomery Scott\authorrefmark{3} and
%Eldon Tyrell\authorrefmark{4}}
%\authorblockA{\authorrefmark{1}School of Electrical and Computer Engineering\\
%Georgia Institute of Technology,
%Atlanta, Georgia 30332--0250\\ Email: mshell@ece.gatech.edu}
%\authorblockA{\authorrefmark{2}Twentieth Century Fox, Springfield, USA\\
%Email: homer@thesimpsons.com}
%\authorblockA{\authorrefmark{3}Starfleet Academy, San Francisco, California 96678-2391\\
%Telephone: (800) 555--1212, Fax: (888) 555--1212}
%\authorblockA{\authorrefmark{4}Tyrell Inc., 123 Replicant Street, Los Angeles, California 90210--4321}}

% use only for invited papers
%\specialpapernotice{(Invited Paper)}

% make the title area
\maketitle

\begin{abstract}
We have quantified a short term instability budget for an optical frequency standard based on cold, freely expanding calcium atoms.  Such systems are the subject of renewed interest due to their high frequency stability and relative technical simplicity compared to trapped atom optical clocks.  By filtering the clock laser light at 657~nm through a high finesse cavity, we observe a slight reduction in the optical Dick effect caused by aliased local oscillator noise.  The ultimately limiting technical noise is measured using a technique that does not rely on a second clock or fs-comb.
\end{abstract}

% no keywords

% For peer review papers, you can put extra information on the cover
% page as needed:
% \begin{center} \bfseries EDICS Category: 3-BBND \end{center}
%
% for peerreview papers, inserts a page break and creates the second title.
% Will be ignored for other modes.
\IEEEpeerreviewmaketitle

\section{Introduction}
Optical frequency standards based on trapped atoms and ions offer unmatched absolute uncertainties~\cite{katori2011optical,chou2010frequency}.  However, a clock based on cold but untrapped atoms may still offer competitive levels of short term stability while remaining technically simpler and more readily commercialized.  Here we discuss limits to the stability of an optical standard featuring cold, freely expanding, Ca atoms~\cite{oates1999diode}.

Briefly, atoms are collected in a magneto-optical trap (MOT) by use of the strong $^{1}S_0 \! \leftrightarrow \! {}^{1}P_1$ transition at 423~nm.  Though this MOT period lasts only 2.8~ms, many atoms are retrapped between clock cycles, yielding a steady-state atom number of order $10^6$.  An external cavity diode laser at 657~nm, stabilized to a ULE glass Fabry-Perot cavity (linewidth $\sim 9$~kHz), is made resonant with the allowed inter-combination transition $^{1}S_0 \! \leftrightarrow \! {}^{3}P_1$.  The MOT is shuttered for 0.44~ms, and a sequence of 657~nm pulses forms a Ramsey-Bord\'{e} atom interferometer~\cite{borde1984optical} in the freely falling and expanding atoms.  Counter-propagating, horizontally aligned pairs of Ramsey pulses suppress residual first-order Doppler shifts, enabling high resolution, and high S/N, spectroscopy~\cite{wilpers2006improved,degenhardt2005calcium}.  Resonant 423~nm light is used to measure the atoms' ground state population before and after the clock pulse sequence. Dividing these collected fluorescence signals yields a spectroscopic feature normalized against atom number fluctuations with a linewidth (FWHM) of $\sim 1$~kHz, and signal-to-noise ratio (S/N) of about 50 in 3.3~ms.  The clock instability limit implied by these results is about $2 \times 10^{-15} (\tau/\text{1s})^{-1/2}$;  which is roughly consistent with the quantum projection noise limit given a coherent fringe contrast of $\sim 10\%$.

We measure the Ca clock instability to be consistently $3-4 \times 10^{-15} (\tau / 1\text{s})^{-1/2}$ when compared against a superior local optical standard~\cite{jiang2011making} via a fs-comb.  We calculate that the optical Dick effect~\cite{quessada2003dick,dick1987local} contributes $3.6 \times 10^{-15} (\tau / 1\text{s})^{-1/2}$ to this instability.  About half of the Dick effect is eliminated by filtering the clock laser through a high-finesse optical cavity to remove high frequency local oscillator noise.  We measure a comparable amount of instability, $1.8 \times 10^{-15} (\tau / 1\text{s})^{-1/2}$, due to technical sources.  In this paper we discuss these contributions and highlight measurement techniques used to quantify them with minimal use of a second clock or fs-comb.
\subsection{A simplified apparatus}
\begin{figure}
\centering
\includegraphics[width=3in]{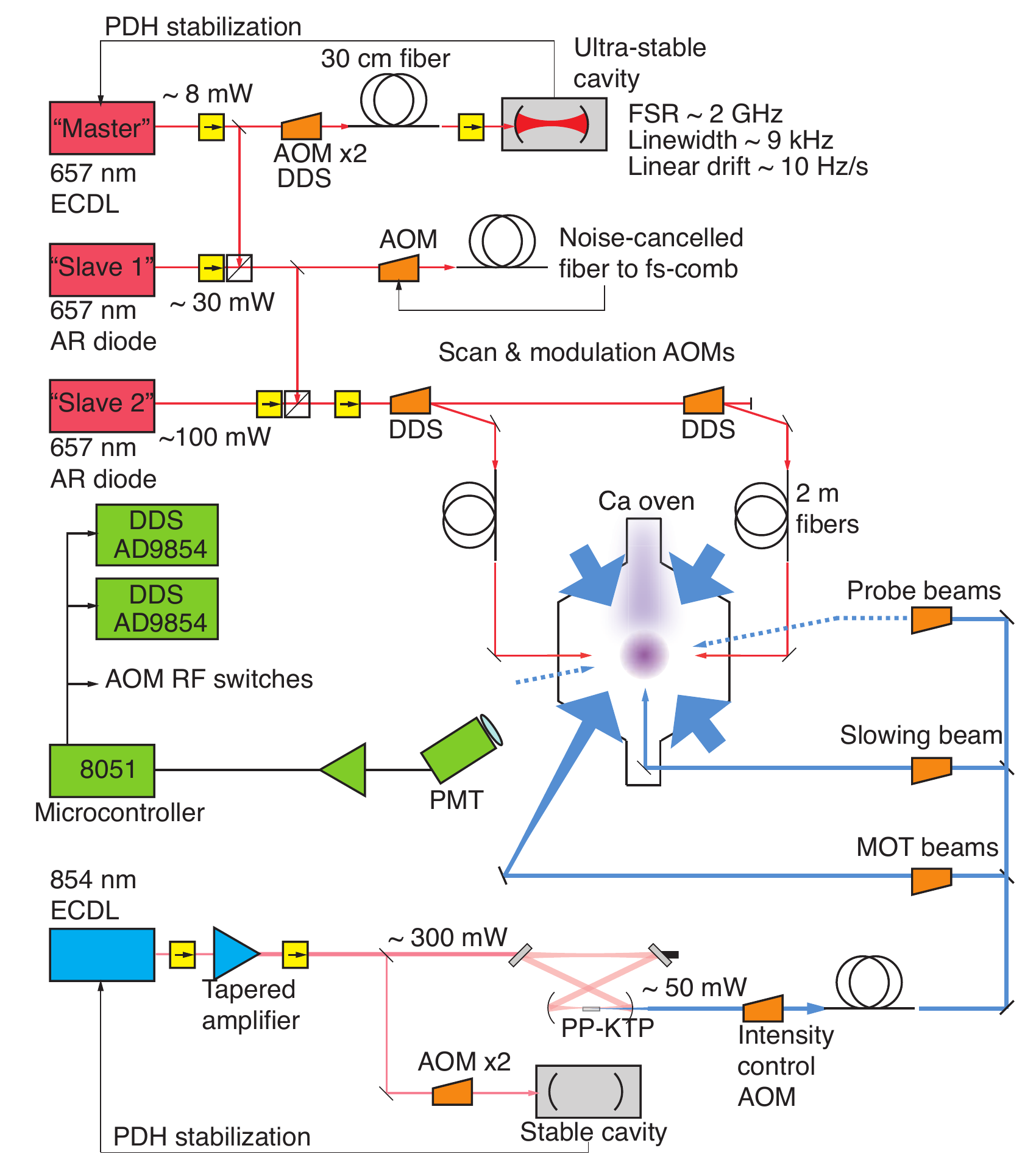}
\caption{The free Ca clock is among the simplest optical frequency standards.  Slowing, cooling, trapping, and probing radiation at 423~nm is derived from a frequency-doubled external cavity diode laser.  A diode laser at 657~nm, AR-coated and placed in a Littman external cavity, is stabilized to an ultra-high finesse optical cavity.  Injection-locked `slave' laser systems boost the 657~nm power to enable very short Ramsey pulses and homogenous broadening of the clock transition up to the residual Doppler linewidth of $\sim 2$~MHz.  A microcontroller triggers the clock sequence every 3.3~ms, acquires fluorescence data, and servos the clock laser on the atomic resonance.}
\label{fig:apparatus}
\end{figure}
Figure~\ref{fig:apparatus} shows the current version of the apparatus, which has been optimized for stability and fast cycle rate, rather than accuracy.  For instance, the MOT gradient magnetic field remains on permanently, eliminating vibrations that result from rapid toggling~\cite{wilpers2007absolute}.  We forgo a quenched second-stage of atom cooling~\cite{wilpers2007absolute} in order to minimize the sequence duration and maximize the atomic interrogation duty cycle.  The effusive atomic beam source is located about 15~cm away from the interaction region;  no Zeeman slower or chirped slowing beam is employed.  Key acousto-optic modulators between the clock laser and the atoms, and between the laser and the stabilization cavity, are driven with maser referenced DDS synthesizers (labeled DDS in Figure~\ref{fig:apparatus}).  The only fiber optics path that is actively noise-cancelled is the long run to the fs-comb.

\section{Local oscillator noise}
\subsection{The optical Dick effect}
\begin{figure}
\centering
\includegraphics[width=3in]{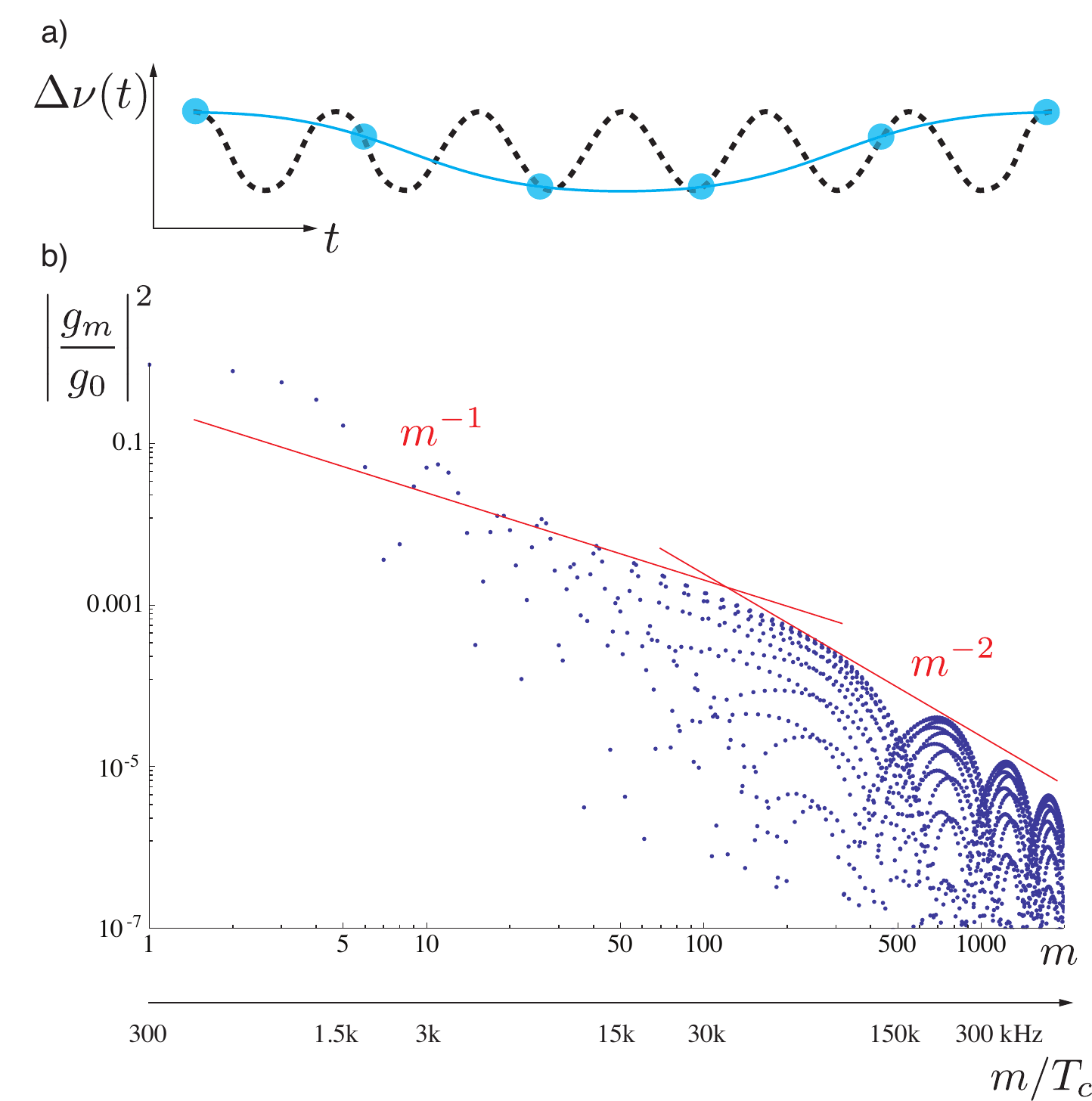}
\caption{(a) A cartoon illustrates the optical Dick effect as an aliasing, or downsampling of noise.  The dashed line represents a particular Fourier component of laser frequency noise close to the clock measurement repetition rate;  the width of each blue circle represents the atomic interrogation period of each clock cycle.  Since these interrogation windows are short, it takes many clock cycles to properly average over the noise component.  The aliased noise (blue line) is not genuine; if written by a servo onto the local oscillator it can actually increase the short term instability of a clock.   (b)  Given the Ramsey-Bord\'{e} pulse sequence parameters, we calculate sensitivity coefficients $g_m/g_0$ in Eq.~\ref{eq:dickEffect} which describe how a local oscillator frequency noise spectrum is aliased into white noise.}
\label{fig:dickEffect}
\end{figure}
Since the clock's repetition rate is 300~Hz, a digital servo locking the laser to the atomic resonance should strongly suppress local oscillator noise at 1~s averaging times. However, since the fraction of the clock sequence spent interrogating the atomic resonance---the duty cycle---is small, high frequency fluctuations are not averaged to zero but instead are down-sampled.  This process, known as the optical Dick effect, results in extra white noise with a fractional instability:
\begin{equation} \label{eq:dickEffect}
\sigma^2_d (\tau) = \frac{1}{\tau} \sum_{m = 1}^\infty S_y(m/T_c) \left| \frac{g_m}{g_0} \right|^2.
\end{equation}
Local oscillator noise $S_y$ at harmonics $m$ of the cycle rate $1/T_c$ are aliased by the imperfect interrogation.  Sensitivity coefficients,
\begin{equation}
g_m \equiv \frac{1}{T_c} \int_0^{T_c} g(t) e^{-2 \pi i m t/T_c} \, dt, \quad (m > 0),
\end{equation}
tend to shrink as the interrogation duty cycle approaches 100\% because the sensitivity of the atomic population to oscillator phase $g(t) = 2 \left. \delta P / \delta \phi \right|_{\delta_\phi \to 0}$ approaches a constant value.  The sensitivity function $g(t)$ for the Ramsey-Bord\'{e} pulse sequence is derived elsewhere~\cite{quessada2003dick}.

Figure~\ref{fig:dickEffect} illustrates the aliasing effect and shows calculated Dick effect coefficients for our nominal clock cycle having an interrogation duty cycle of 13\%.  The Ramsey $\pi/2$ pulses last $T_R = 0.75 \mu$s.  The duration between Ramsey pulses is $T = 215 \mu$s.  The dead-time between pairs of opposing Ramsey pulses is $T_\nu = 5.8 \mu$s.  Increasing the two Ramsey dark periods lasting $T$ would increase resolution and the interrogation duty cycle, but would also inflict a loss of signal;  the lifetime of the excited clock state is 0.4~ms.  Numerical analysis of Eq.~\ref{eq:dickEffect} with our estimated laser noise spectrum $S_y$ shows that more than 99\% of the Dick effect instability is due to harmonics $m < 100$, or laser noise frequency components $f < 30$~kHz.

\subsection{Clock laser filtering}
\begin{figure}
\centering
\includegraphics[width=3in]{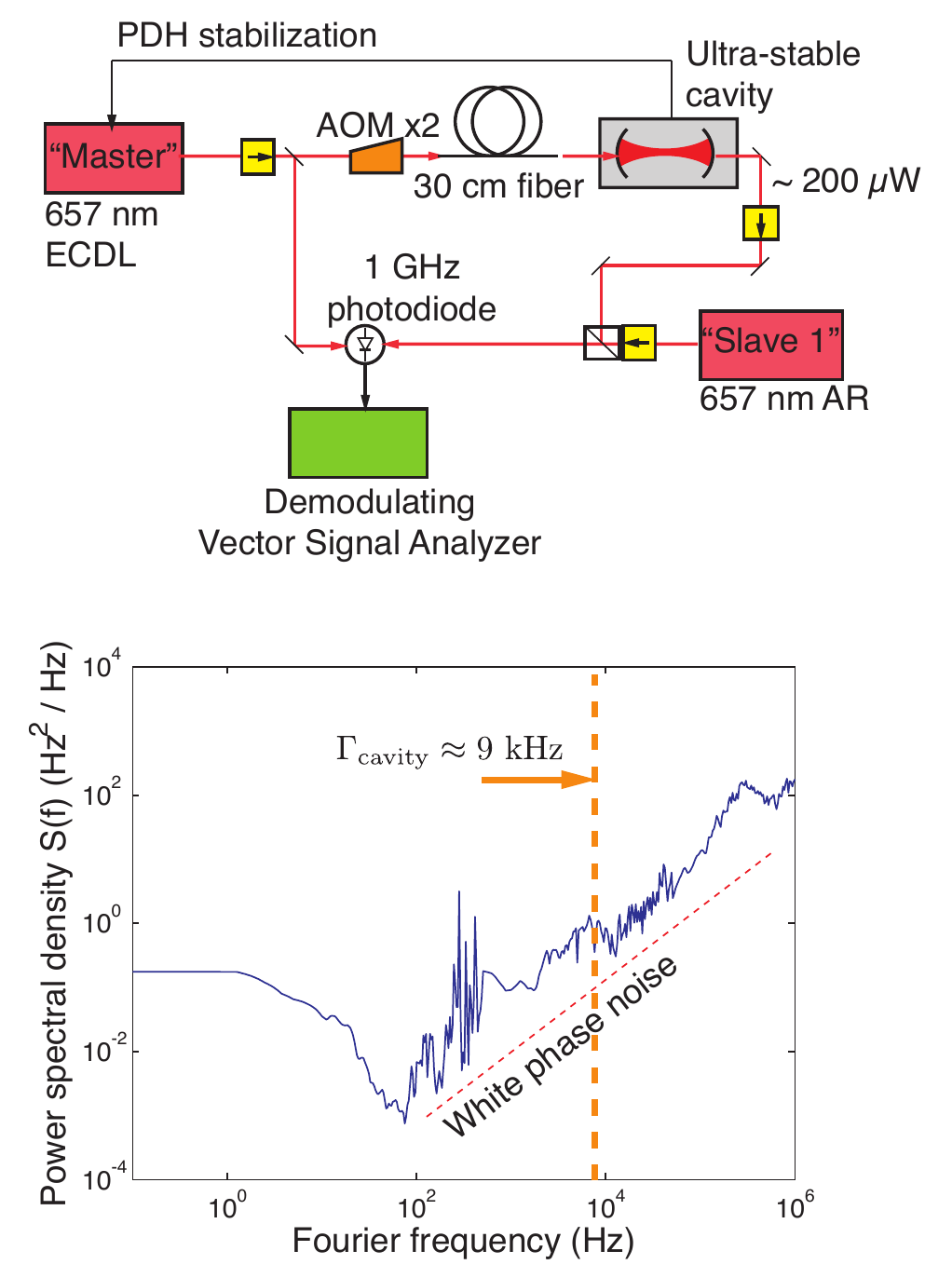}
\caption{Fast laser frequency noise is suppressed by filtering through a high-finesse optical cavity;  the cavity acts to average out fluctuations faster than the photon ring-down timescale of $\tau = 1/\Gamma_\text{cavity} \sim 0.1$~ms.  To determine the transfer function of the cavity, we heterodyne laser beams before and after transmission through a high finesse optical cavity used for laser stabilization.  The noise added by the injection locking to a second laser is independently evaluated.}
\label{fig:phaseNoiseHet}
\end{figure}
We find the instability of the cavity-stabilized clock laser (while unlocked to the atomic resonance) to be $2.5 \times 10^{-15}$ after 1 second of averaging by comparison to a superior optical reference.  This instability is consistent with that derived from the laser's noise spectrum from 1~Hz--1~kHz found by demodulating the laser's beatnote with a referenced fs-comb.  Faster laser noise components, perhaps due to residual phase noise not suppressed by the Pound-Drever-Hall stabilization scheme, can be downsampled by the Dick effect.  To examine what fast noise might be suppressed by filtering the clock laser, we heterodyne clock laser beams before and after transmission through the high-finesse stabilization cavity.  As shown in Figure~\ref{fig:phaseNoiseHet},  noise rising as $f^2$ is observed that contributes directly very little to the local oscillator instability at 1 second;  however, through the Dick effect mechanism, the noise above $\Gamma_\text{cavity} \approx 9$~kHz is responsible for roughly half of the Dick effect instability. The unfiltered and filtered laser Dick effect instability limits are
\begin{align*}
\sigma_\text{d,} (\tau) &= 3.6 \times 10^{-15} (\tau/\text{1s})^{-1/2} \\
\sigma_\text{d, filt.} (\tau) &= 2.4 \times 10^{-15} (\tau/\text{1s})^{-1/2}
\end{align*}
We do not see as large an improvement from clock laser filtering as other groups~\cite{nazarova2008low} due to our shorter cycle time and larger interrogation duty cycle.

\subsection{Other noise sources}
\begin{figure}
\centering
\includegraphics[width=3in]{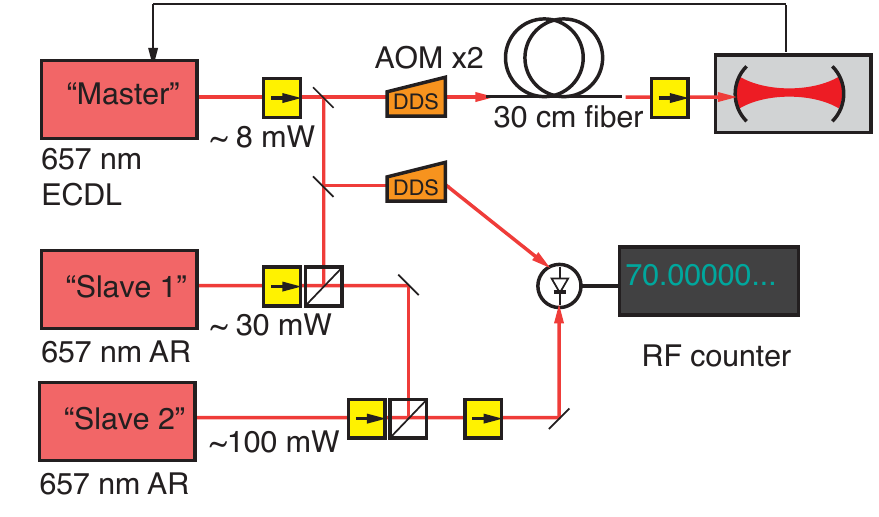}
\caption{Illustrated here is a heterodyne between the master external cavity diode laser and an injection-locked `slave' laser, which is itself seeded by another `slave' laser.  An RF beatnote is created by shifting one of the beams with an acousto-optic modulator (AOM) driven by a maser-referenced synthesizer (labeled DDS).  The Allan deviation of maser-referenced RF counter is related to the optical frequency fractional Allan devation relevant to Ca clock operation with $\sigma_y(\tau) = \sigma_\nu(\tau) / f_\text{optical}$. Direct counting of a DDS frequency yields a baseline of $\sigma_y (1\text{s}) = 6.0 \times 10^{-18}$.  Table~\ref{tab:heterodynes} shows the results for several such heterodyne measurements;  all instabilities are much lower than the observed clock instability.  The noise spectra have not been studied in detail.}
\label{fig:hetExample}
\end{figure}
\begin{table}
% increase table row spacing, adjust to taste
\renewcommand{\arraystretch}{1.3}
\caption{Instability contributions determined by heterodyne measurements.}
\label{tab:heterodynes}
\begin{center}
% Some packages, such as MDW tools, offer better commands for making tables
% than the plain LaTeX2e tabular which is used here.
\begin{tabular}{ll} \hline
Heterodyne test & $\sigma_y(\tau = \text{1s})$ \\ \hline
DDS counting baseline	& $6.0 \times 10^{-18}$ \\
Before/after cavity & $6.3 \times 10^{-16}$ \\
Before/after 1.5m PM fiber & $1.3 \times  10^{-16}$ \\
Master vs.\ injection locked `slave 1' & $1.6 \times 10^{-16}$ \\
Master vs.\ injection locked `slave 2' & $1.8 \times 10^{-16}$
\end{tabular}
\end{center}
\end{table}
We performed heterodyne measurements between the master ECDL beam and several derived laser beams---after injection locking, after short fiber optic runs, after acousto-optic modulation (AOM)---to identify other sources of noise that might contribute to the clock instability. An example arrangement is shown in Figure~\ref{fig:hetExample}.  In each heterodyne experiment, AOM frequencies and the RF counter were maser referenced.  Table~\ref{tab:heterodynes} expresses these noise contributions as fractional Allan deviations of the optical frequency at 1 second.

\section{Technical and total noise evaluation}
\subsection{Self-evaluation of instability}
\begin{figure}
\centering
\includegraphics[width=3in]{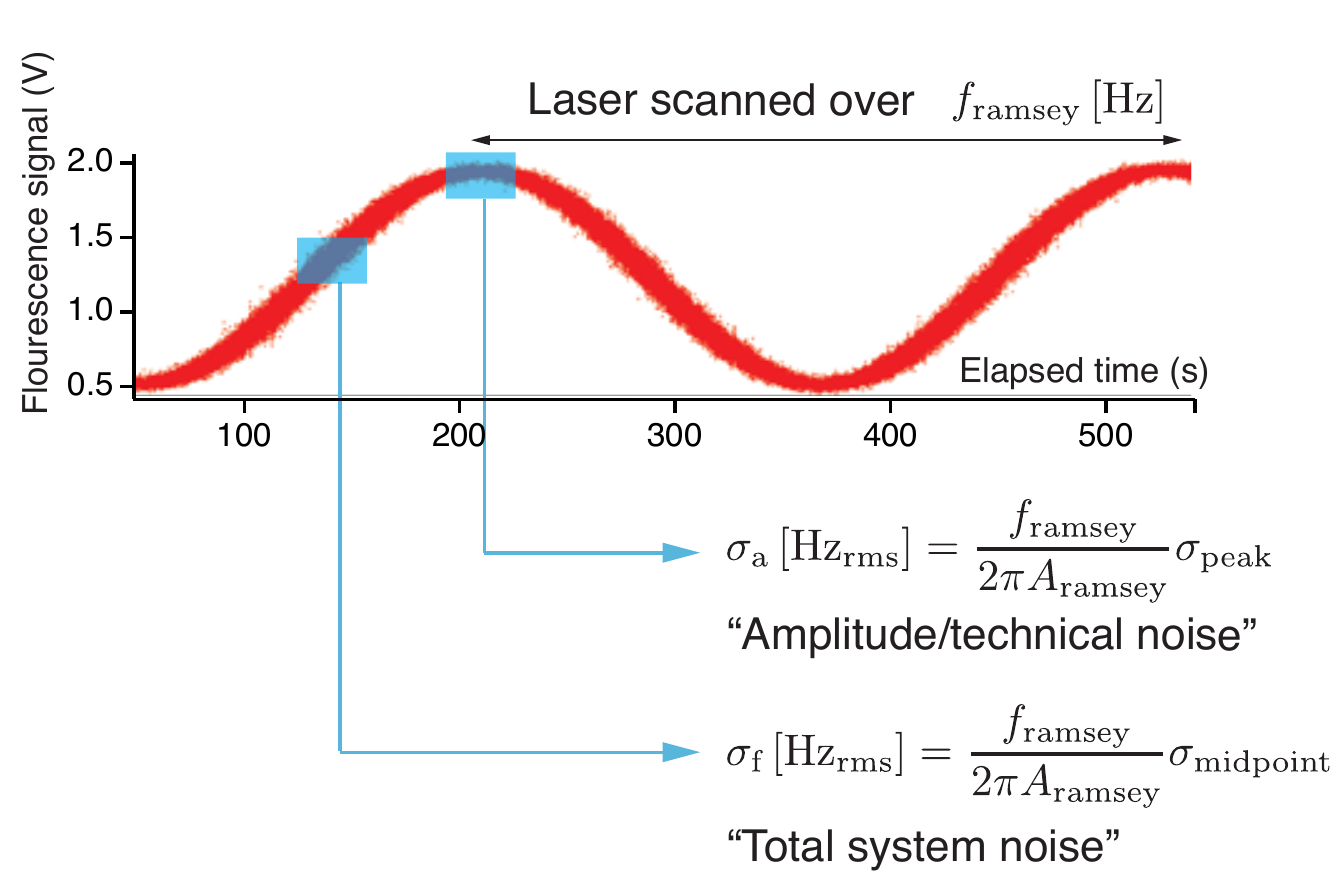}
\caption{In the absence of a second local oscillator or atomic reference, one can still derive useful absolute instability of a clock by using the atomic reference as a frequency discriminator.  Here, the unlocked clock laser is slowly scanned over a Ramsey-Bord\'{e} fringe.  Point-to-point rms deviations in the fluorescence signal are larger near the frequency-sensitive nodes.  Technical noise, which is presumed to be largely laser frequency independent, is observable at the peaks and troughs of the fringe.  The clock's operational parameters can be optimized to minimize each noise level.  The technical noise level sets a limit on ultimate system stability even in the absence of local oscillator noise.  The noise level at the fringe nodes is quantitatively consistent with the overall clock performance when compared against superior optical standards.}
\label{fig:totalSystemNoise}
\end{figure}
Without the benefit of a second, independent clock laser or atomic reference, one can still evaluate some aspects of the absolute instability of a Ca clock apparatus.  With a clock laser detuning $\nu$, frequency fluctuations $\delta \nu(t)$, and Ramsey dark period $T$, the spectroscopy method produces a fluorescence signal \cite{borde1984optical}:
\begin{equation}
\rho_e(t) \propto A \cos \left(2 \pi \times 4T \times (\nu + \delta \nu(t) \right) + \delta \rho_e(t).
\end{equation}
The laser detuning $\nu$ can be slowly scanned over this feature;  laser noise $\delta \nu (t)$ is translated into fluctuations in fluorescence proportional to the local slope $\partial \rho_e / \partial \nu$.  At the Ramsey fringe nodes, this slope is constant and computable using the Ramsey fringe amplitude $A$ and separation $f_\text{Ramsey} = 1/(4T)$. The root-mean-square deviation of successive measurements of $\rho_e$ can be expressed in as a frequency instability by
\begin{equation}
\sigma_\nu (\tau = T_c) [\text{Hz}] = \frac{1}{2\pi A (4T)} \sigma_\rho [V_\text{rms}].
\end{equation}
At the peaks and troughs of the Ramsey fringe we observe technical noise $\delta \rho_e(t)$ which is largely independent of the clock laser frequency but nonetheless contributes noise when the servo is locked to a nodal point.  At these lock-points, the technical noise will be equivalent to uncorrelated frequency noise using the same conversion factor $1/({2\pi A 4T)}$.  Several potential sources of technical noise are suppressed by normalizing the signal to atom number, and by dithering the Ramsey-Bord\'{e} fringe pattern to remove the incoherent excitation background.  The remaining technical noise could include shot-to-shot atom number fluctuations imperfectly accounted for by the normalization pulse, fluctuations in 657~nm pulse power, and frequency noise on the 423~nm probe pulses.  Path length, pointing, and wavefront fluctuations between Ramsey pulses in a single clock cycle also appear as excess technical noise, but are dependent on the laser frequency as well.  In general, fluctuations in any systematic effect leading to inaccuracy adds instability as well, though dithering of the Ramsey fringe suppresses such effects varying slower than half the cycle rate.

Fourier transforms of recorded $\rho_e(t)$ series at the peaks, toughs, and nodes indicate these noises are largely white from 0.5---150 Hz, and thus have equivalent noise spectra of $S_\nu (f) = \sigma_\nu^2 / (f_\text{rep})$, where $f_\text{rep}$ is the clock cycle rate.  Effective fractional Allan deviations can be constructed with $\sigma_y^2(\tau) = (S_\nu/2) (1/\tau) (1/f_\text{optical})$.

\begin{figure}
\centering
\includegraphics[width=3in]{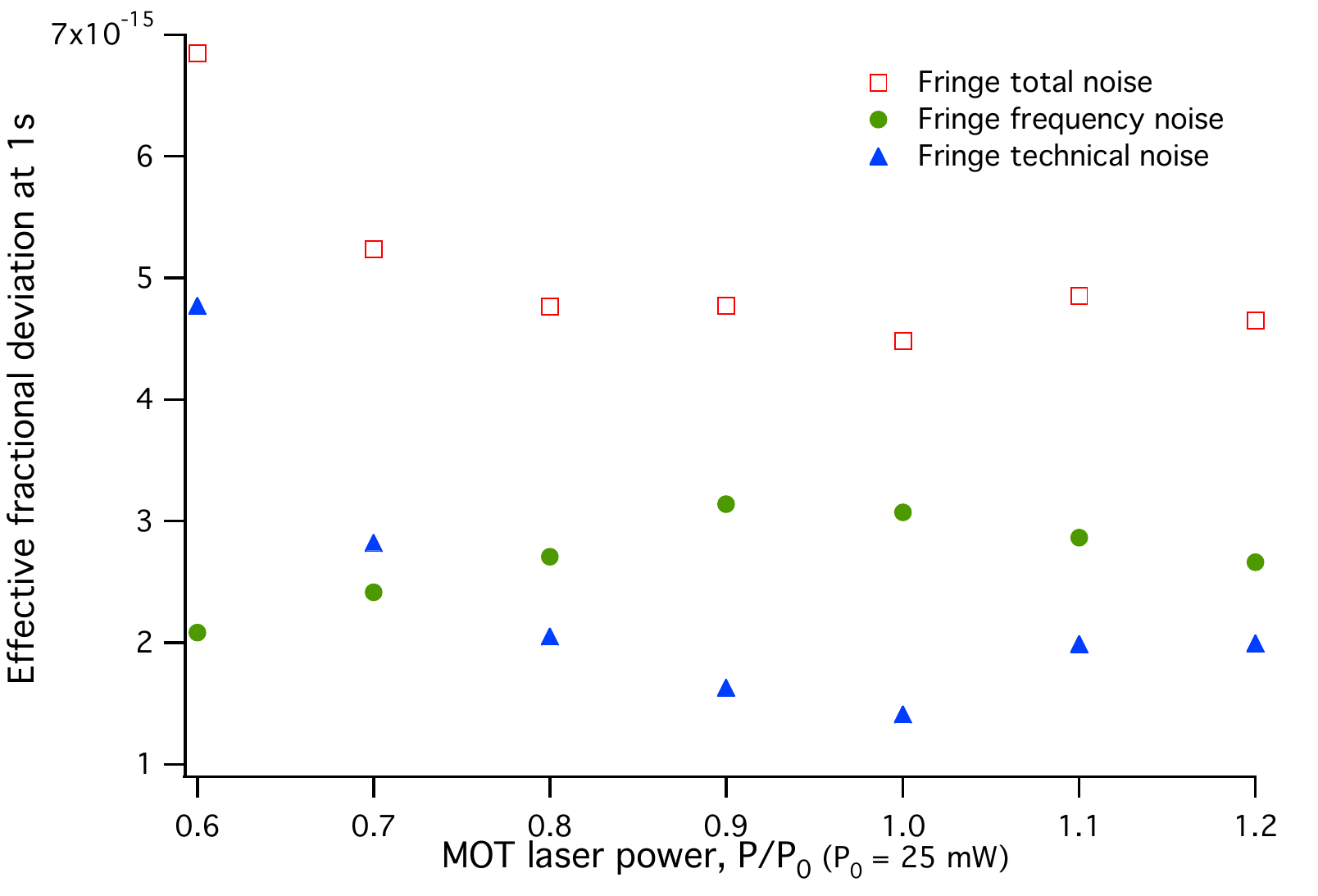}
\caption{The instability of the clock system is measured by using the atomic resonance as a frequency discriminator in open loop operation.  Noise on the Ramsey fringe peaks and nodes can be related to effective Allan deviations due to technical and frequency-dependent noise, respectively.  Here we see how these noises vary as a function of MOT trapping power.  An optimum operation condition, the minimum of the total noise, is found without using a second clock or fs-comb.}
\label{fig:noiseVsMOTpower}
\end{figure}

Figure~\ref{fig:noiseVsMOTpower} demonstrates the utility of this technique.  The MOT laser power is varied around its nominal power of 25~mW, and the effective deviation at 1s implied by the open-loop fringe noise is plotted.  Technical and frequency noise (corresponding to peak and node noise levels) are separately plotted, and their sum agrees well with the observed clock instability when the unlocked local oscillator noise is subtracted.

\begin{table*}
% increase table row spacing, adjust to taste
\renewcommand{\arraystretch}{1.3}
\caption{A short-term instability budget for a free calcium optical clock.}
\label{tab:budget}
\begin{center}
% Some packages, such as MDW tools, offer better commands for making tables
% than the plain LaTeX2e tabular which is used here.
\begin{tabular}{m{1in}cm{2in}m{1in}} \hline
Noise source & Symbol & Measurement technique & Fractional Allan devision at 1 s \\ \hline
\raggedright Local oscillator, unlocked & $\sigma_L$& \raggedright Noise spectrum measured from beat with Yb-clock on fs-comb & $2.5 \times 10^{-15}$ \\ \hline
\raggedright Dick effect, unfiltered & $\sigma_\text{d}$ & Calculated & $3.6 \times 10^{-15}$ \\
\raggedright Dick effect, filtered through cavity & $\sigma_\text{d,filt.}$ & \raggedright Filtered noise obtained by filtered/unfiltered heterodyne & $2.4 \times 10^{-15}$ \\
\raggedright Technical noise, and atom/photon shot noise & $\sigma_\text{tech.}$ &`Self-evaluated' open-loop Ramsey fringe noise at peaks/troughs & $1.8 \times 10^{-15}$ \\ 
\raggedright Additional noise from injection locking, fibers, etc. & $\sigma_\text{add.}$ & Heterodyne measurements & $0.6 \times 10^{-15}$ \\ \hline
\multicolumn{3}{l}{Quadrature sum of $\sigma_\text{d, filt.}, \sigma_\text{tech.}, \sigma_\text{add.}$ (filtered laser case)} & $ 3.0 \times 10^{-15}$ \\
\multicolumn{3}{l}{Quadrature sum of $\sigma_\text{d}, \sigma_\text{tech.}, \sigma_\text{add.}$ (unfiltered laser case)} & $4.0 \times 10^{-15}$ \\
\multicolumn{3}{l}{Measured instability against Yb lattice clock, filtered laser, typical} & 3.5(5) $\times 10^{-15}$ \\ \hline
\multicolumn{3}{l}{`Self-evaluated' open loop Ramsey fringe noise at nodes, $\sigma_L$ subtracted in quadrature} & $3.6 \times 10^{-15}$ \\ \hline
\end{tabular}
\end{center}
\end{table*}

\section{Conclusion}
Previous efforts have aggressively pursued an uncertainty budget for the Ca optical frequency standard. In Table~\ref{tab:budget}, we present instead an instability budget at 1s of averaging with no account of ultimate accuracy.  We conclude that three different methods of evaluating the instability share good agreement:  direct measurement against a superior optical standard, a quadrature sum of the calculated optical Dick effect instability and known technical noise sources, and an open loop measurement using the atoms as a frequency discriminator.  The latter technique never relies on a second clock or fs-comb, but it is directly sensitive to the open-loop local oscillator noise.

Another important result is that while filtering the clock laser through our high-finesse cavity decreases the Dick effect limit somewhat, it remains comparable to the technical noise that would ultimately limit the system if the local oscillator were significantly improved.  Further study must focus on identifying and reducing the principle sources of technical noise.  Since this level is thought to be near that of quantum projection noise, an increase in the number of atoms might be necessary for further improvements.

\section*{Acknowledgment}
% optional entry into table of contents (if used)
%\addcontentsline{toc}{section}{Acknowledgment}
The authors would like to thank colleagues T.\ Fortier and S.\ Diddams for fs-comb measurements, and A.\ Ludlow, N.\ Lemke, R. Fox, and Y.\ Jiang for comparison measurements with the Yb optical lattice clock and local oscillator.  We thank R.\ Fox, A.\ Curtis, F.\ Bondu, G.\ Wilpers, and Y.\ LeCoq for their contributions to earlier versions of the Ca apparatus.

% trigger a \newpage just before the given reference
% number - used to balance the columns on the last page
% adjust value as needed - may need to be readjusted if
% the document is modified later
%\IEEEtriggeratref{8}
% The "triggered" command can be changed if desired:
%\IEEEtriggercmd{\enlargethispage{-5in}}

% references section
% NOTE: BibTeX documentation can be easily obtained at:
% http://www.ctan.org/tex-archive/biblio/bibtex/contrib/doc/

% can use a bibliography generated by BibTeX as a .bbl file
% standard IEEE bibliography style from:
% http://www.ctan.org/tex-archive/macros/latex/contrib/supported/IEEEtran/bibtex
%\bibliographystyle{IEEEtran.bst}
% argument is your BibTeX string definitions and bibliography database(s)
%\bibliography{IEEEabrv,../bib/paper}
%
% <OR> manually copy in the resultant .bbl file
% set second argument of \begin to the number of references
% (used to reserve space for the reference number labels box)
%\begin{thebibliography}{1}
%\end{thebibliography}

\bibliographystyle{IEEEtran} 
\bibliography{IEEEabrv,shermanCaBib}

% that's all folks
\end{document}